\documentclass[twocolumn,nofootinbib,preprintnumbers]{revtex4}
\usepackage{pslatex}
\usepackage[pdftex]{graphicx}

\usepackage{psfrag}
\usepackage{epsfig}
\usepackage{color}
\usepackage{cancel}
\usepackage{amsmath}
\usepackage{enumitem,xspace}

\def\refe@jnl#1{{#1}}
\def\aj{\refe@jnl{Astron.~J.}}
\def\araa{\refe@jnl{Annu.~Rev.~Astron.~Astrophys.}}
\def\apj{\refe@jnl{Astrophys.~J.}}
\def\apjl{\refe@jnl{Astrophys.~J.~Lett.}}
\def\aap{\refe@jnl{Astron.~Astrophys.}}
\def\mnras{\refe@jnl{Mon.~Not.~R.~Astron.~Soc.}}
\def\prd{\refe@jnl{Phys.~Rev.~D}}
\def\fcp{\refe@jnl{Fund.~Cos.~Phys.}}
\def\physrep{\refe@jnl{Phys.~Rep.}}
\def\physlett{\refe@jnl{Phys.~Lett.}}

\def\invisible#1{  }

\def\lsim{\mathrel{\lower4pt\hbox{$\sim$}} 
\hskip-9.5pt\raise1.6pt\hbox{$<$}\;} 
 \def\gsim{\mathrel{\lower4pt\hbox{$\sim$}} 
\hskip-9.5pt\raise1.6pt\hbox{$>$}\;}

\newcommand{\fermi}{{\sc Fermi-LAT}\xspace}
\newcommand{\hess}{{\sc HESS}\xspace} 
\newcommand{\amstwo}{{\sc AMS-02}\xspace} 
\newcommand{\pamela}{{\sc PAMELA}\xspace}

\definecolor{orange}{rgb}{1,0.5,0}

\begin{document}

\title{Effective Theory of Dark Matter Decay into Monochromatic Photons and its Implications:
\\ Constraints from Associated Cosmic-Ray Emission}

\author{Michael Gustafsson, Thomas Hambye and Tiziana Scarn\`a}
\email{mgustafs@ulb.ac.be;thambye@ulb.ac.be;tscarna@ulb.ac.be}
\affiliation{Service de Physique Th\'eorique\\
 Universit\'e Libre de Bruxelles\\ 
 Boulevard du Triomphe, CP225, 1050 Brussels, Belgium}


\begin{abstract}
We show that there exists only a quite limited number of higher dimensional operators which can naturally lead to a slow decay of dark matter particles into monochromatic photons. As each of these operators inevitably induces decays into particles other than photons, we show that the $\gamma$-lines it induces are always accompanied by a continuum flux of cosmic rays. Hence constraints on cosmic-ray fluxes imply constraints on the intensity of $\gamma$-lines and {\it vice versa}. A comparison with up to date observational bounds shows the possibilities to observe or exclude cosmic rays associated to $\gamma$-line emission, so that one could better determine the properties of the DM particle, possibly discriminating between some of the operators.
\end{abstract}

 \maketitle
\section{Introduction}
\label{sec:Introduction}
One of the best ``smoking-gun'' signals for establishing the existence of an annihilating or decaying dark matter (DM) particle is the possible observation of a cosmic  $\gamma$-ray line \cite{Bergstrom:1988fp}.  Forthcoming satellites \cite{Galper:2012ji,Maestro:2013gul,DAMPE} and air Cherenkov telescopes \cite{Bringmann:2011ye,Bergstrom:2012vd,Doro:2012xx}, but also  current instruments like the Fermi large area telescope (\fermi) \cite{Ackermann:2012qk,FermiSymposium} and the \hess instrument \cite{Abramowski:2013ax,HESSII}, will allow to probe this possibility with further sensitivity.
From DM particle annihilations the amount of monochromatic $\gamma$-rays produced is expected to be limited (even if in some cases it can saturate the present observational bounds) because it is in general loop suppressed with respect to the total annihilation cross section, that is 
generically constrained by the DM relic density. For a decay, on the contrary, the amount of monochromatic rays emitted could  {\it a priori} be much larger. Even if in many scenarios the DM particle is not expected to decay at all, there exists a well-motivated theoretical framework where DM would naturally decay with a lifetime larger than the age of the Universe: if its stability is due to an accidental low energy symmetry that has no reason to be respected by any ultraviolet (UV) theory, just as expected for the proton in the standard model (SM).
In this case the effect of the UV physics causing the decay is suppressed by powers of the UV scale. At low energy, this can be parametrized in full generality by writing down the most general effective theory respecting the low energy symmetries of the model. 
For a DM particle mass around the electroweak scale (as supported by thermal scenarios), it turns out that a decay suppressed by 4 powers of the grand unified theory (GUT) scale  $ \Lambda_{GUT} \simeq 10^{15}$ GeV, {\it i.e.} induced by dimension 6 operators, leads to a lifetime which can give fluxes of cosmic rays (CRs) of the order of current observational sensitivities, $\tau \simeq \Lambda_{GUT}^4/M_{DM}^5  \simeq 10^{26}$~s  \cite{GUTdecay,Hambye:2008bq}.
This is a too nice opportunity to probe the GUT scale to not study it in more detail. 
For an explicit example of a setup with an accidental symmetry, which can lead to DM decay into $\gamma$-lines through dimension 6 operators, see reference \cite{Hambye:2008bq}.

The use of an effective theory picture for a DM decay is fully justified because, unlike an annihilation, it necessarily requires a large UV scale (unless one would invoke extremely tiny coefficients). As explained below, this model-independent gauge invariant effective operator approach implies that when a DM particle decays, it does so to several final states.
It is therefore different from the usual model independent approach of considering indirect detection constraints on separated single decay final states, as {\it e.g.}~Refs.~\cite{noteffectivedecay}. The amount of CRs that these several final states induce is also 
in general larger than those from electroweak corrections on a  $\gamma$-line final state, as considered in Ref.~\cite{Ciafaloni:2010ti}. By allowing a systematic determination of the $\gamma$-ray line emission possibilities, and by properly taking care of the low energy symmetries of the theory, the low energy effective operator language is the appropriate one to approach a series of important related questions: What are the theoretical expectations to see a $\gamma$-line from a DM decay (including, what effective structure a UV theory must induce to do so)?\footnote{For this question it is important to keep in mind that the radiative operators we will list below could be induced either directly from the UV physics, or from low energy loop correction to other UV induced DM decay operators.} What are the possibilities to see a $\gamma$-line, given the CR constraints, and conversely to  see CRs associated to the observation of a $\gamma$-line? And from there, what are the possibilities to discriminate among the various effective radiative operators, hence to have information on the DM particle whose decay would have produced the $\gamma$-line, and maybe on the UV physics associated?

\section{Full list of radiative operators}
\label{sec:Full list}
To write down the most general effective theory of DM decays is a quite tremendous task, especially if one allows for new particles into which the DM particle could decay.
In scenarios of accidentally stable DM, such particles can be expected, see {\it e.g.}~Ref.~\cite{Hambye:2008bq}. \
However, for the production of monochromatic lines it turns out that this can be done, {\it i.e.}~the number of operators (up to dimension 6) is quite limited.

This is due to the stringent criteria an operator must fulfill in this case:
 
\begin{enumerate}[topsep=4pt, partopsep=2pt,itemsep=8pt,parsep=2pt,leftmargin=1.3 em]
  \item[a)] First, obviously the operator must contain the DM field. 
  \item[b)] Secondly,  in order for the operator to lead to a two-body decay with a photon, it must not contain too many fields (except eventually scalars that could be replaced by their vacuum expectation values (vev)) and it must contain either a hypercharge $F_Y^{\mu\nu}$ or a $SU(2)_L$ $F_L^{\mu\nu}$ field strength. If the DM particle is neutral, as we will assume here, the photon cannot come from a covariant derivative because in the two body decays all fields are necessarily neutral. One could eventually have a photon emitted from an operator that contains a new $U(1)$ gauge field that kinematically mix with the hypercharge gauge boson, ${\cal L} \owns \varepsilon F_Y^{\mu\nu}F'_{\mu\nu}$ (rendering the various neutral particles to be effectively milli-charged). We will not consider this possibility here. It gives a range of bounds on the emission of $\gamma$-ray lines which is similar to the one obtained below (see Ref.~\cite{GHTfuture} for details). 
  \item[c)] Thirdly, some operators can be related to other ones through various relations, equations of motions, shift of a derivative or use of the fact that the commutator of two covariant derivatives $D_\mu$, $D_\nu$ gives $F_{\mu\nu}$. However one must be careful in using these relations. The criteria we apply here is that through these relations an operator  can be dropped from the list only if in this way there is a one-to-one correspondence between this operator and another one already in the list. Otherwise both operators must be kept because in general they give different 
ratios of $\gamma$-line to CRs.
 \end{enumerate}

Applying the criteria above there are only three possible general dimension 5 structures, one for each of the three  types of DM particle we consider here, scalar, fermion or vector:  $\phi_{DM} F_{\mu\nu} F^{\mu\nu}$,
$\bar{\psi} \sigma_{\mu\nu} \psi_{DM} F^{\mu \nu}$, $F_{\mu\nu}^{DM} F^{\mu\nu} \phi$ respectively. 
By specifying the nature of the $F^{\mu\nu}$ field strengths one obtains 9 possible operators, 5 for a scalar candidate, and 2 each for a fermion and a vector candidate
\begin{alignat}{7}
&{\cal O}_{\phi_{DM}}^{(5)YY}	&\;\equiv\;\;\;& \phi_{DM} F_{Y\mu\nu} F^{\mu\nu}_Y  	\;\;\quad&&\phi_{DM}=(1,0)		\;\;\quad&& A
\label{OphiDM5YY}\\
&{\cal O}_{\phi_{DM}}^{(5)YL}	&\;\equiv\;\;\;& \phi_{DM} F_{L\mu\nu} F^{\mu\nu}_Y   	\;\;\quad&&\phi_{DM}=(3,0) 		\;\;\quad&& B
\label{OphiDM5YL}\\
&{\cal O}_{\phi_{DM}}^{(5)LL}	&\;\equiv\;\;\;& \phi_{DM} F_{L\mu\nu} F^{\mu\nu}_L      	\;\;\quad&&\phi_{DM}=(1/3/5,0)		\;\;\quad&& D_{m}
\label{OphiDM5LL}\\
&{\cal O}_{\phi_{DM}}^{(5)YY'}	&\;\equiv\;\;\;& \phi_{DM} F_{Y\mu\nu} F^{\mu\nu}_{Y'}   		\;\;\quad&&\phi_{DM}=(1,0)		\;\;\quad&& A_x
\label{OphiDM5YY'}\\
&{\cal O}_{\phi_{DM}}^{(5)LY'}	&\;\equiv\;\;\;& \phi_{DM} F_{L\mu\nu} F^{\mu\nu}_{Y'}    	\;\;\quad&&\phi_{DM}=(3,0) 		\;\;\quad&& C_x
\label{OphiDM5LY'}\\[5pt]
&{\cal O}_{\psi_{DM}}^{(5)Y} 	&\;\equiv\;\;\;& \bar{\psi} \sigma_{\mu\nu} \psi_{DM} F^{\mu \nu}_Y  \;\;\quad&&\psi_{DM}\cdot \psi=(1,0)	\;\;\quad&& A_x
\label{OpsiDM5Y}\\
&{\cal O}_{\psi_{DM}}^{(5)L} 	&\;\equiv\;\;\;& \bar{\psi} \sigma_{\mu\nu} \psi_{DM} F^{\mu \nu}_L    \;\;\quad&&\psi_{DM}\cdot \psi=(3,0) 	\;\;\quad&& C_{x,m}
\label{OpsiDM5L}\\[5pt]
&{\cal O}_{V_{DM}}^{(5)Y}	&\;\equiv\;\;\;& F_{\mu\nu}^{DM} F^{\mu\nu}_Y \phi  	 	\;\;\quad&&\phi=(1,0)  			\;\;\quad&& A_x
\label{OVDM5Y}\\
&{\cal O}_{V_{DM}}^{(5)L}		&\;\equiv\;\;\;& F_{\mu\nu}^{DM} F^{\mu\nu}_L \phi  		 \;\;\quad&&\phi=(3,0)			\;\;\quad&& E_x
\label{OVDM5L}
\end{alignat} 
where $\phi_{DM}/ \psi_{DM}$ denotes the multiplet whose neutral component $\phi^0_{DM}/ \psi^0_{DM}$ is the DM particle.  By "$(n,Y)$'' we specify what must be the size $n$ of the $SU(2)_L$ multiplets and their hypercharge $Y$. $F'^{\mu\nu}$ stands for a new possible low energy gauge field and the vector DM operator $F_{\mu\nu}^{DM}$ stands for 
an abelian or non-abelian DM field strength (in practice it will not be necessary to make this distinction in the following). $\psi$ and $\phi$ are meant to be either SM fields when allowed by gauge invariance or new low energy fields. The symbols $A-E_{x,m,v}$ stand for a classification of the operators' possible astrophysical signals,  and will be explained in Sec.~\ref{sec:lineVScont}.

As for the dimension 6 operators the number of  possibilities is also remarkably limited. Two general structures are singled out for the scalar case and three for the fermion and vector cases, leading to 7 scalar operators
\begin{eqnarray}
{\cal O}_{\phi_{DM}}^{1YY}&\equiv& \phi_{DM} F_{Y\mu\nu} F^{\mu\nu}_Y \phi     \quad\,\,\,\, \phi_{DM}\cdot \phi=(1,0) \quad\quad \,\,\, A
\label{OphiDMYY}\\
{\cal O}_{\phi_{DM}}^{1YL}&\equiv& \phi_{DM} F_{L\mu\nu} F^{\mu\nu}_Y\phi     \quad \,\,\,\, \phi_{DM}\cdot \phi=(3,0)\quad \quad \,\,\, B
\label{OphiDMYL}\\
{\cal O}_{\phi_{DM}}^{1LL}&\equiv& \phi_{DM} F_{L\mu\nu} F^{\mu\nu}_L\phi    \quad \,\,\,\, \phi_{DM}\cdot \phi=(1/3/5,0)\,\,\, C_{x,m}\,\,\,
\label{OphiDMLL}\\
{\cal O}_{\phi_{DM}}^{1YY'}&\equiv& \phi_{DM} F_{Y\mu\nu} F^{\mu\nu}_{Y'}\phi    \quad\,\,\,\, \phi_{DM}\cdot \phi=(1,0)\quad \quad \,\,\,A_x
\label{OphiDMYY'}\\
{\cal O}_{\phi_{DM}}^{1LY'}&\equiv& \phi_{DM} F_{L\mu\nu} F^{\mu\nu}_{Y'}\phi  \quad\,\,\,\, \phi_{DM}\cdot \phi=(3,0)\quad \quad \;\;C_x
\label{OphiDMLY'}\\
{\cal O}_{\phi_{DM}}^{2Y}&\equiv& D_\mu \phi_{DM} D_\nu \phi F^{\mu\nu}_Y    \quad \phi_{DM}\cdot \phi=(1,0) \quad\;\;\,A_{x,m,v}
\label{OphiDMY}\\
{\cal O}_{\phi_{DM}}^{2L}&\equiv& D_\mu \phi_{DM} D_\nu \phi F^{\mu\nu}_L   \quad \phi_{DM}\cdot \phi=(3,0)   \quad\;\;\,C_{x,m,v}
\label{OphiDML}
\end{eqnarray}
 to 6 fermion operators
\begin{eqnarray}
{\cal O}_{\psi_{DM}}^{1Y}&\equiv& \bar{\psi} \sigma_{\mu\nu} \psi_{DM} F^{\mu \nu}_Y \phi    \,  \quad \bar{\psi}\cdot \psi_{DM}  \cdot \phi=(1,0)\,\,\,\,A_{x,m}\,\,
\label{Opsi1DMY}\\
{\cal O}_{\psi_{DM}}^{1L}&\equiv& \bar{\psi} \sigma_{\mu\nu} \psi_{DM} F^{\mu \nu}_L \phi     \,  \quad \bar{\psi}\cdot \psi_{DM} \cdot \phi=(3,0)\,\,\,\,C_{x,m}\,\,
\label{Opsi1DML}\\
{\cal O}_{\psi_{DM}}^{2Y}&\equiv& D_\mu\bar{\psi} \gamma_\nu  \psi_{DM} F^{\mu \nu}_Y         \quad \bar{\psi}\cdot \psi_{DM}=(1,0) \quad\,\,\,\,\,A_x\,\,
\label{Opsi2DMY}\\
{\cal O}_{\psi_{DM}}^{2L}&\equiv& D_\mu\bar{\psi} \gamma_\nu  \psi_{DM} F^{\mu \nu}_L         \quad \bar{\psi}\cdot \psi_{DM}=(3,0)\quad\,\,\,\,C_{x,m}\,\,
\label{Opsi2DML}\\
{\cal O}_{\psi_{DM}}^{3Y}&\equiv& \bar{\psi} \gamma_\mu D_\nu \psi_{DM} F^{\mu \nu}_Y         \quad \bar{\psi}\cdot \psi_{DM}=(1,0)\quad\,\,\,\,A_x\,\,
\label{Opsi3DMY}\\
{\cal O}_{\psi_{DM}}^{3L}&\equiv& \bar{\psi} \gamma_\mu D_\nu \psi_{DM} F^{\mu \nu}_L         \quad \bar{\psi}\cdot \psi_{DM}=(3,0)\quad\,\,\,C_{x,m}\,\,
\label{Opsi3DML}
\end{eqnarray}
and to 5 vector operators
\begin{eqnarray}
{\cal O}_{V_{DM}}^1&\equiv& F_{\mu\nu}^{DM} F^{\mu\rho}_Y F^\nu_{Y'\rho}       \qquad\qquad\qquad\qquad\;\;\;\; A_x
\label{OV1DMY}\\
{\cal O}_{V_{DM}}^{2Y}&\equiv& F_{\mu\nu}^{DM} F^{\mu\nu}_Y \phi \phi'     \,\,\,\quad \quad\quad \phi \cdot \phi'=(1,0) \,\,\,\,\,\,A_x
\label{OV2DMY}\\
{\cal O}_{V_{DM}}^{2L}&\equiv& F_{\mu\nu}^{DM} F^{\mu\nu}_L \phi \phi'        \,\,\,\quad \quad\quad \phi \cdot \phi'=(3,0) \,\,\,\,\,\, D_{x,m}
\label{OV2DML}\\
{\cal O}_{V_{DM}}^{3YY'}&\equiv& D_\mu^{DM}\phi D_\nu^{DM} \phi' F^{\mu\nu}_Y       \,\,\,  \quad \phi \cdot \phi'=(1,0) \,\,\,\,\,\,A_{x,m}\label{OV3YY'}
\label{OV3DMY}\\
{\cal O}_{V_{DM}}^{3LY'}&\equiv& D_\mu^{DM}\phi D_\nu^{DM} \phi' F^{\mu\nu}_L        \, \,\,  \quad \phi \cdot \phi'=(3,0) \,\,\,\,\,\,D_{x,m}.
\label{OV3DML}
\end{eqnarray}
By $D_\mu^{DM}$ we mean a covariant derivative that contains the DM vector field.\footnote{In Ref.~\cite{Hambye:2008bq} an explicit example can be found of an accidental symmetry setup leading to the operators of Eqs.~(\ref{OV2DMY}) and (\ref{OV3DMY}). Note also that in Ref.~\cite{Garny:2010eg} there are examples of heavy scalar and heavy vector setups whose exchange induces dimension 6 four-fermion interactions that at one loop induce a $\psi_{DM}\rightarrow \gamma \nu$ decay. The effective amplitude for this process is the same as the ones that the dimension 5 operators of Eq.~(\ref{OpsiDM5Y}) or Eq.~(\ref{OpsiDM5L}) give. This exemplifies the fact, 
to keep in mind,  that dimension 5 operators for a decay can naturally have a ``dimension 6 suppression'' of the lifetime.}

Along the operator criteria defined above, note that one must still add to the list three types of operators, whose structures are somehow more involved.

First, a few operators that contain two covariant derivatives on a same field. There  are two structures of this kind for the scalar case, 
$D_\mu D_\nu \phi_{DM} \phi F^{\mu\nu}$,
$ \phi_{DM}  D_\mu D_\nu\phi F^{\mu\nu} $ 
and one for the vector case,
$D_\mu^{DM} D_\nu^{DM}\phi \phi' F^{\mu\nu}$ (with $F^{\mu\nu}$ the hypercharge or $SU(2)_L$ field strength). 
These operators are reducible to a combination of operators where the two covariant derivatives are replaced by field strengths. 

Secondly, there are four fermionic operators with a ``$\slash\hspace{-2.2mm}D$'', $\bar{\psi}_{DM} \sigma_{\mu\nu}\, {\slash \hspace{-2.2mm}D} \psi F_{Y,L}^{\mu\nu}$ and  $\bar{\psi}_{DM}  \slash\hspace{-2.5mm}\overleftarrow{D}\sigma_{\mu\nu}  \psi F_{Y,L}^{\mu\nu}$. Through the equation of motion of the spinors, these
operators are reducible to the same operator with $\slash\hspace{-2.2mm}D$ replaced by a mass or a scalar field or a combination of several of these operators.

Thirdly, there are also operators with dual fields. The full list of operators with dual fields can be obtained by replacing one field strength in the operators of the list above by its dual one. Note nevertheless in this case that Eqs.~(\ref{OpsiDM5Y})-(\ref{OpsiDM5L}) and (\ref{Opsi1DMY})-(\ref{Opsi1DML}) do not lead to any new operator (since they give back the same operator with the non-dual field) and Eqs.~(\ref{OphiDMY})-(\ref{OphiDML}) do not give any $\gamma$-line. Also Eqs.~(\ref{Opsi2DMY})-(\ref{Opsi3DML}) give operators that are reducible to a combination of the same operator without dual and of the operator of Eq.~(\ref{OpsiDM5Y}) or (\ref{OpsiDM5L}).

\medskip
In the following we will consider the phenomenology of operators of Eqs.~(\ref{OphiDM5YY}--\ref{OV3DML}) individually, as usually done in effective theories. Few comments on the possible effects of considering linear combination of several operators can be found in Sec. \ref{sec:final comments}. 
To discriminate operators, or sets of operators, it is not sufficient to observe a $\gamma$-line at a given energy and with a given intensity, because these observables can be accounted for by any operator, by adjusting the DM mass and the $1/\Lambda$ or $1/\Lambda^2$ coefficient in front of the dimension 5 or 6 operators. Additional information is needed, either on the $\gamma$-line spectra they give (Sec.\ref{sec:multiple lines}),  or on the associated continuum spectrum of CRs (Sec. \ref{sec:lineVScont}-\ref{sec:SM}).

Note that in the following we will not consider explicitly the phenomenology of operators above that are combinations of the operators  of Eqs.~(\ref{OphiDM5YY}-\ref{OV3DML}), {\it i.e.}~of operators with 2 covariant derivatives on the same field, of operators with a $\slash\hspace{-2.2mm}D$ and of the operators of Eqs.~(\ref{Opsi2DMY})-(\ref{Opsi3DML}) with dual fields. As for all other dual field operators, they give the same phenomenology than their non-dual partner.


\section{Single or multiple $\gamma$-lines?}
\label{sec:multiple lines}

\begin{figure}[t]
\includegraphics[width=0.49 \columnwidth]{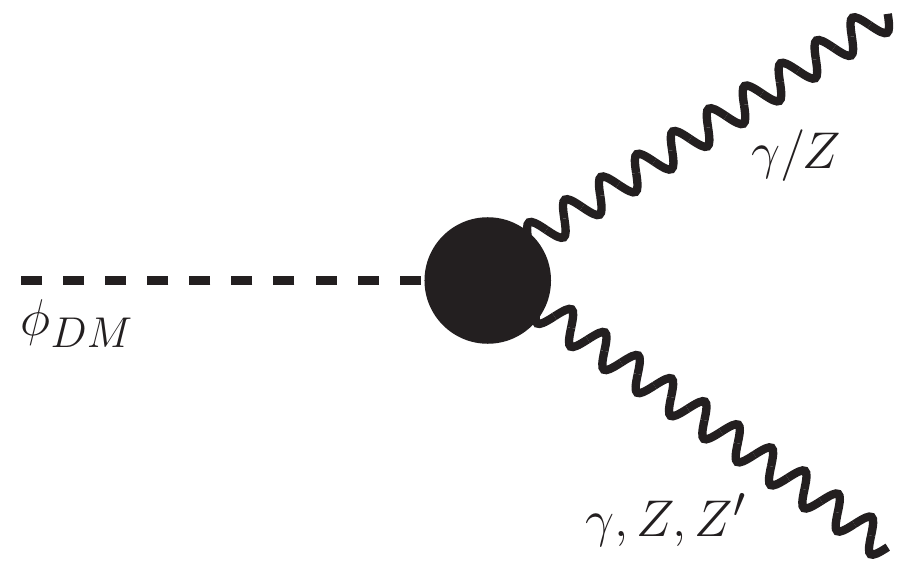}
\includegraphics[width=0.49\columnwidth]{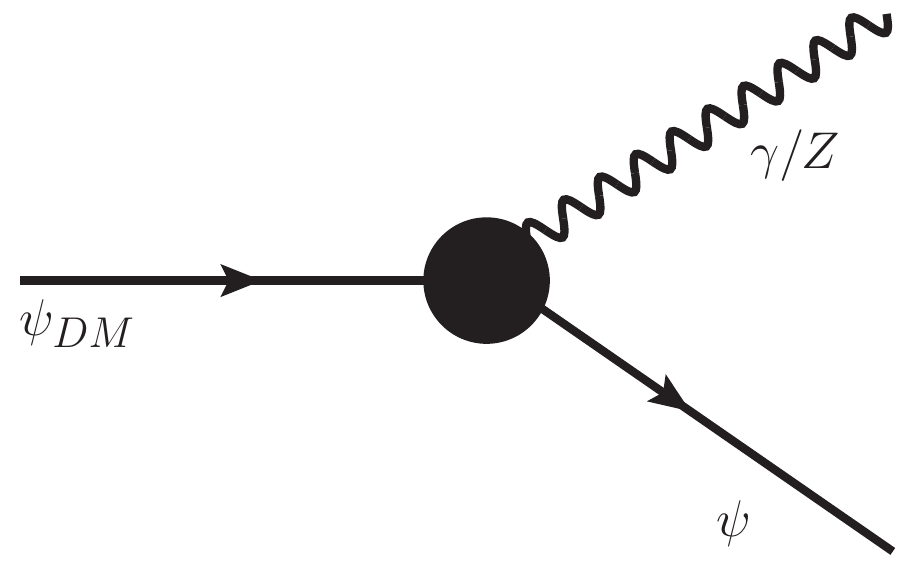}
\includegraphics[width=0.49 \columnwidth]{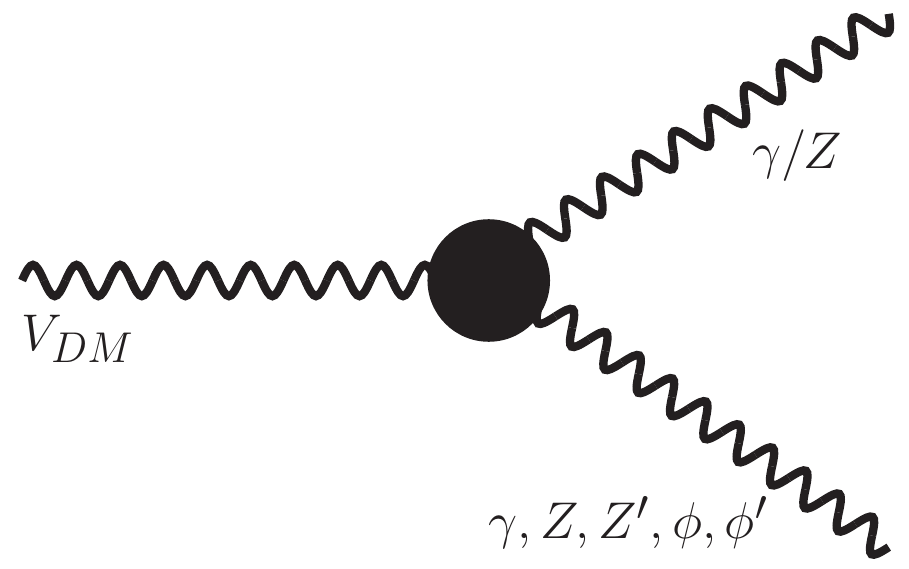}
\caption{
The DM decays into monochromatic photons and their associated final-state particles as induced by the effective operators. }
 \label{fig:operator}
\end{figure}

Let us start by discussing a simple way to, in principle, distinguish between some operators in the above list: from the different number of $\gamma$-lines they can give and the relative strength and energy of these $\gamma$-lines. Fig.~\ref{fig:operator} shows all types of two-body final states including $\gamma$-lines that Eqs.~(\ref{OphiDM5YY})-(\ref{OV3DML}) give. The operators giving more than one $\gamma$-line can be separated into two categories, depending on if they include one or two SM field strengths. 

On the one hand all the operators with two SM field strengths in the operator give peaks associated to the $\gamma\gamma$ and $\gamma Z$ channels. This could be the case for six operators, \mbox{Eqs.~(1-3, 10-12)}. The relative strength of the two $\gamma$-lines are  
\begin{equation}
\Gamma_{\gamma\gamma} /\Gamma_{\gamma Z}=\frac{\cos^2\theta_W}{2 \sin^2\theta_W} F,\;\; \frac{\sin^2 2\theta_W}{2 \cos^22\theta_W} F,\;\;\frac{\sin^2\theta_W}{2 \cos^2\theta_W} F \label{eq:line_Z}
\end{equation}
for the operators including $F_{Y\mu \nu} F_Y^{\mu \nu},  F_{Y\mu \nu}F_L^{\mu \nu}$ and $ F_{L \mu \nu}F_L^{\mu \nu}$ respectively, with 
$F={m^6_{DM}}/{(m^2_{DM}-m^2_Z)^3}$ and $\theta_W$ the Weinberg angle.
If $m_{DM}$ is above but not too far above the $m_Z$ scale, these three types of operators could thus be distinguished from each other because both peak strengths could be separately resolved at energies equal to $m_{DM}/2$ and $m_{DM}/2(1-m^2_Z/m^2_{DM})$). In practice only one $\gamma$-line will appear when the detector resolution and sensitivity are not good enough to separate too close monochromatic lines \cite{Bergstrom:2012vd}.

On the other hand there exists also 6 operators with only one SM field strength which could also give  two $\gamma$-lines: the scalar DM ones in  Eqs.~(\ref{OphiDMY})-(\ref{OphiDML}) which allow a $\gamma Z$ and $\gamma Z'$ decay (with a $Z'$ from the covariant derivatives), the vector DM of Eqs.~(\ref{OV2DMY})-(\ref{OV2DML})  and  Eqs.~(\ref{OV3DMY})-(\ref{OV3DML})  which if $\phi\neq\phi'$ allow both a $\gamma \phi$ and a $\gamma \phi'$ channel. If the mass of the $Z'$, $\phi$ or $\phi'$ differs sizably from $m_Z$ these operators would typically give different energy gaps between the $\gamma$-lines compared to the operators with two SM field strengths. 
In particular for these operators, if these masses are not many times smaller than $m_{DM}$, and have sufficiently small decay widths, one could have an observable multiple peak structure also for $m_{DM} \gg m_Z$; at $m_{DM}/2(1-m^2_Z/m^2_{DM}) \simeq m_{DM}/2$ energy and at $m_{DM}/2(1-m^2_{Z',\phi,\phi'}/m^2_{DM})$ energies.
That would single out these 6 operators. Information on the height and energy of both peaks provides further informations on the masses and couplings involved in the effective operators.
For Eqs.~(\ref{OphiDMY}) and (\ref{OphiDML}) one has 
\begin{equation}
\Gamma_{\gamma Z}/\Gamma_{\gamma Z'}
=  \frac{g^2T_3^2}{\cos^2\theta_W g^2_{Y'}Y'^2} \Bigg(\frac{m^2_{DM}-m^2_{Z}}{m^2_{DM}-m^2_{Z'}}\Bigg)^3 \label{eq:line_Z'}
\end{equation}
with  $g$  the $SU(2)_L$  coupling and $ g_{Y'}, Y'$ are the extra gauge group coupling and charge of the scalar field under this gauge group. 
For Eqs.~(\ref{OV2DMY})-(\ref{OV3DML}) one has
\begin{equation}
\Gamma_{\gamma \phi }/\Gamma_{\gamma \phi'}= 
\frac{\left\langle \phi'\right\rangle^2}{\left\langle \phi\right\rangle^2} \Bigg(\frac{m^2_{DM}-m^2_{Z}}{m^2_{DM}-m^2_{Z'}}\Bigg)^3 \label{eq:line_phi}
\end{equation}
with $\left\langle \phi\right\rangle$ and $\left\langle \phi'\right\rangle$  the $\phi, \phi'$ vevs. 
None of these 6 operators give any $\gamma\gamma$ final state.

The remaining 15 operators, {\it i.e.}\ all the fermion DM particles' operators Eqs.~(\ref{OpsiDM5Y}--\ref{OpsiDM5L},\ref{Opsi1DMY}--\ref{Opsi3DML}), and the scalar and vector DM particles' operators in (\ref{OphiDM5YY'}--\ref{OphiDM5LY'},\ref{OVDM5Y}--\ref{OVDM5L},\ref{OphiDMYY'}--\ref{OphiDMLY'}, \ref{OV1DMY}), give only a single $\gamma$-ray line, and would thus be excluded as an explanation to the observation of multiple monochromatic $\gamma$-lines. Of course, a DM particle can always give multiple lines by appearing in several operators involving fields with different masses.

\section{Continuum rays associated to a $\gamma$ line} \label{sec:lineVScont}

In the previous section we have seen how multiple $\gamma$ lines might allow to single out some operators. We will now focus on a feature shared by all operators in our list, which will allow us to characterize them further: all allowed operators that give a decay into a $\gamma$-ray line will unavoidably also produce, with different amounts, decays into particles producing CRs. This is a consequence of the fact that  gauge symmetries dictate the possible effective operator structures. It is particularly clear that, if kinematically allowed, a $Z$ boson is always among the final states since the hypercharge and $SU(2)_L$ neutral gauge eigenstates are made of a mixture of a photon and a Z 
boson.\footnote{This also holds for a DM annihilation scenario when the $\gamma$-line(s) is induced by effective operators \cite{Rajaraman:2012db
} or induced at one loop in explicit models \cite{Bergstrom:1994mg
}. This property is also characteristic of specific gravitino decay scenarios \cite{Covi:2008jy}.}

This multiple final state property is the crucial one why, by merely studying the corresponding continuum of CRs at lower energies, we can put upper bounds on the possible strength of the $\gamma$-line signal for any given operator. 
In the following we will focus our analysis on the large $m_{DM}\ge200$ GeV mass region, where the production of CR from a given operator is unavoidable, {\it i.e.} $m_{DM}$ is sufficiently large so that  channels with up to two SM gauge bosons are kinematically fully open. 
We will also make the assumption that the mass of any new beyond SM final state particle is negligible with respect to the DM particle, so that if the operator leads to several peaks, they all show up at the same $E_\gamma \simeq m_{DM}/2$ and their contributions are summed.\footnote{If on the contrary the mass of this new state is close to $m_{DM}$, some channels could be more phase space suppressed than others, modifying the constraints. 
Note also that, if one decreases $m_{DM}$ towards the $m_Z$ threshold, the $R_{\gamma/CR}$ ratio could go to zero or to infinity, depending on whether the operator induces a decay to $\gamma\gamma$.}

Inspection of the full list of operators shows that there is no single operator which can give 
a ratio of monochromatic photons to $Z$ production larger than (what we call prediction A) 
\begin{equation}
A:\,\,\frac{n_\gamma}{n_Z}=\frac{\cos^2 \theta_W}{\sin^2 \theta_W}\quad \hbox{or} \quad R_{\gamma/CR}\equiv \frac{n_\gamma}{n_{CR}}=\frac{\cos^2 \theta_W}{\sin^2 \theta_W n_{CR/Z}},
\label{ngammanZmax}
\end{equation}
where $CR=\bar{p},\gamma_{c}, e^+,...$ stands for any CR which could be produced out of the $Z$, with $n_{CR/Z}$ the number of CRs produced per $Z$ boson at a given energy. In contrast to the monochromatic $\gamma$-line, the $\gamma_c$ stands for the continuum of lower energy photons produced by hadronization and electroweak correction \cite{Cirelli:2010xx} connected to the production of $Z$ bosons.\footnote{Final state radiation and internal Bremsstrahlung processes could mimic a line-like signal. We do not account for these processes in the numerator of the $R_{\gamma/CR}$ ratio.} Given the fact that CR fluxes are observationally bounded from above, this ratio sets the absolute upper bound on the intensity of $\gamma$-ray lines that a single operator can give, to be determined below.

It is out of the scope of this work to list all the exact predictions the various operators give for the monochromatic photons to CR emission (and the corresponding upper bound they give, see Ref.~\cite{GHTfuture}) because a single operator may give different ratios depending on the quantum numbers of the particles it contains, and the nature of the $\phi$ or $\psi$ particles they might involve. However the $A-E_{x,m,v}$ labels we have put beside each operator in Eqs.~(\ref{OphiDM5YY}-\ref{OV3DML}) allow already to see rather well what  is  going on for each of them.
Varying between all possible multiplets in the operators, each operator turns out to give a \emph{maximum} ratio which is given either by A above or by one of the four following relations
\begin{eqnarray}
B:\,\, R_{\gamma/CR}&=&\frac{1}{n_{CR/Z} } \\
C:\,\, R_{\gamma/CR}&=&\frac{\sin^2 \theta_W}{\cos^2 \theta_W n_{CR/Z}}\\
D,E:\,\, R_{\gamma/CR}&=&\frac{\sin^2 \theta_W}{\cos^2 \theta_W n_{CR/Z} + c_{D,E} (n_{CR/W^+}+n_{CR/W^-})}\qquad \label{ngammanZphi1LL}
\end{eqnarray}
with $c_{D,E}=1/4,1$ for D and E, respectively. Signal ratio A  is characteristic of operators containing only $F_Y^{\mu\nu}$, B is typical of those containing both $F_Y^{\mu\nu}$ and  $F_L^{\mu\nu}$,  and signal ratios C to E correspond to operators including only $F_L^{\mu\nu}$. 
These bounds are obtained assuming that the $\phi$ and $\psi$ particles which may show up in the two-body final state do not produce any CRs (as could be the case for a beyond SM particle). 

From these maximal ratios, each operator may give a $R_{\gamma/CR}$ ratio that can be smaller in three possible ways, if the extra $\phi$ or $\psi$ particle does decay to CRs (in these cases we put a ``x'' index label), if the operator gives different ratios for different multiplets (``m'' label), or if the $R_{\gamma/CR}$ depends on the (unknown) vev of a scalar field (``v'' label).  There are also operators which give a totally fixed ratio. In these cases we do not attach any index label. This is the case for two A type operators and for all $B$ cases. 

The ``$x$'' and ``$m$'' variations of $R_{\gamma/CR}$ that can arise for an operator 
are relatively limited. Given the fact that in each case there is already production of CRs from the $Z$ (and sometimes $W$) boson, if we  assume the extra particles ``x'' give an amount of CR similar to what a $Z$ or a $W$ boson produces, these particles cannot modify largely the ratio. Typically, the change is not more than a factor of a few or at most, in rare cases of type A predictions, around one order of magnitude. Similarly, among the different multiplet configurations that do lead to a $\gamma$-line, a given operator can have the $R_{\gamma/CR}$ ratio varying by a factor of a few up to at most a factor $5-6$ (except for Eq.~(\ref{Opsi1DML}) which allows a factor $\sim10$ variation because it involves two fields on top of the DM and gauge boson fields). For example many operators of class C with an index ``$m$'' may easily give, for some multiplets, $W^\pm$ bosons in the final state, in that case the signal ratios can give predictions equal or close to $D$ or $E$. Therefore from these effects it is not possible to go far below the maximum signal ratio, A-E, that each operator are assigned. 
Only by varying the value of the vev of a beyond SM scalar field can we go far below the operators maximal $R_{\gamma/CR}$ ratio value, which is only possible for the two ``v'' labelled operators in Eqs.~(\ref{OphiDMY})-(\ref{OphiDML}). The latter possibility stems from the fact that the $Z+\phi$ production is proportional to $m_Z^2$ whereas the $\gamma$-line productions are instead proportional to the vev of $\phi$, and thus disappear when the vev goes to zero.\footnote{Note that for Eqs.~(\ref{OV2DMY})-(\ref{OV3DML}) there is also a vev dependence but it is much milder.}

As examples of ``v'' specific cases which necessarily give ratios much smaller than the A-E predictions, let us consider $\phi$ being a triplet of hypercharge 2 in the operators of Eq.~(\ref{OphiDMY}) and (\ref{OphiDML}). We call these predictions F$_Y$ and F$_L$ respectively.
For these setups the $\gamma$-line production is proportional to $v_\phi^2$, whereas the $Z \phi$ production is proportional to $m_Z^2$. Since the vev of a triplet is constrained to be small by electroweak data ($|v|\lesssim 4$~GeV), the $R_{\gamma/CR}$ is suppressed by at least a factor $v^2_\phi/m^2_Z\lesssim 10^{-2}$. 
As a result, for example the $F_L$ case of Eq.~(\ref{OphiDML}) lies a factor $\sim 1000$ below the absolute one of case A for the $R_{\gamma/CR}$ ratio (and with a $\gamma$-line intensity so weak that basically it cannot be distinguished from the associated continuum of $\gamma$ rays, see below). This shows that operators that give ``v'' labelled predictions crucially depend on the electroweak quantum numbers of the particles involved and for some multiplets predict very weak lines. As stated above the same operators of Eq.~(\ref{OphiDMY}) and Eq.~(\ref{OphiDML}) can saturate predictions A and C (for instance for $\phi$ a SM singlet, whose vev can be large and break an extra gauge symmetry, so that the DM particle can decay to the associated gauge boson(s) and a $\gamma$).

\section{General results}
\label{sec:General results}

%
\begin{figure}[t]
\center{\includegraphics[width=0.95 \columnwidth]{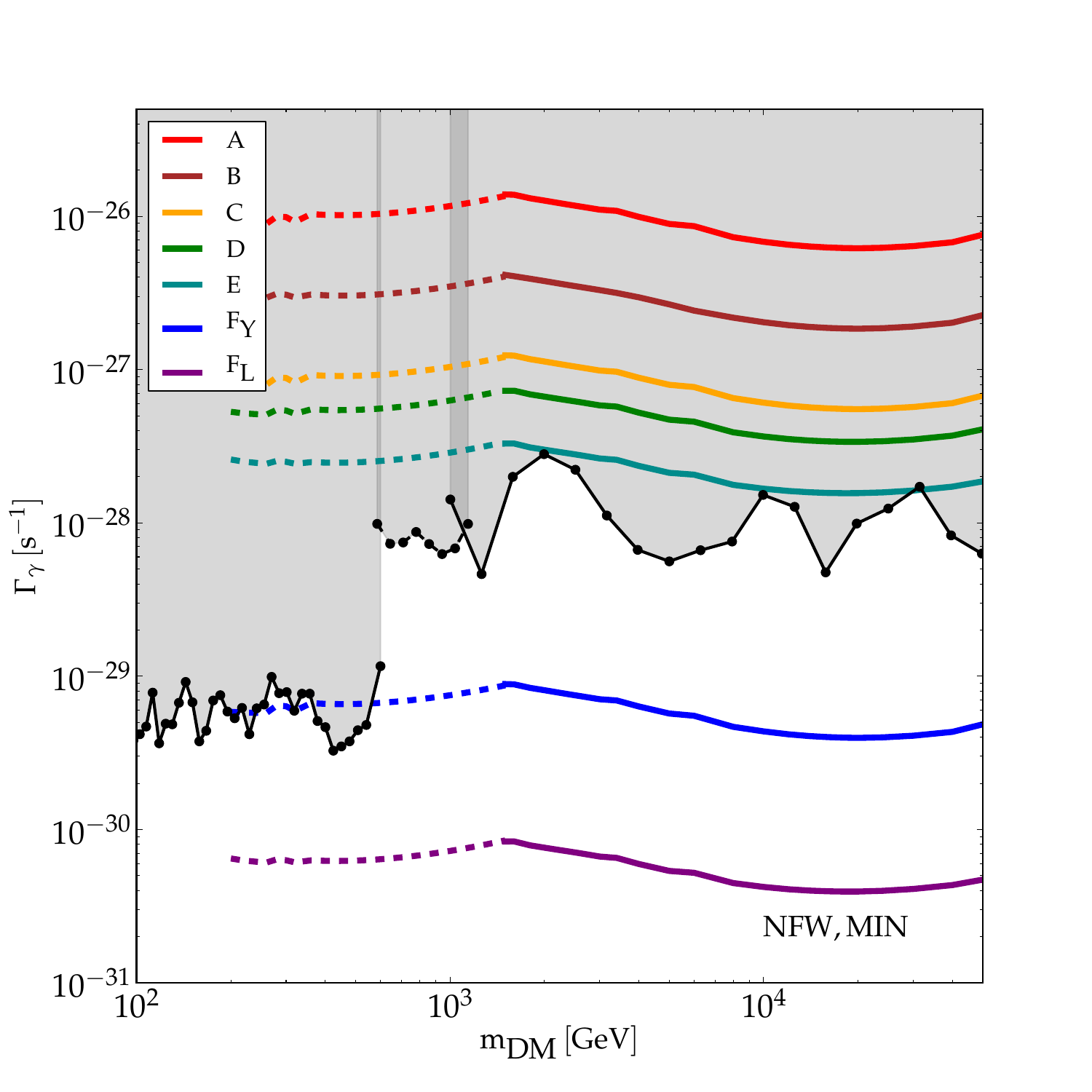}}
\caption{Upper bounds on the decay rate into monochromatic photons for the predicted ratios A-F. 
Dashed curves represent the indirect constraints imposed by antiproton data from \pamela, whereas solid curves are derived from the isotropic diffuse $\gamma$-ray measurement by the \fermi. The curves are, from top to bottom, in the same order as mentioned in the figure's legend. The grey shaded area represents the direct observational exclusion region from \fermi and \hess data.}\label{fig:abcde}
\end{figure}
%

Once the ratio between the predicted line signal and the lower energy continuum of CR is known, 
observational constraints on CRs put indirect limits on the possible strength of any associated $\gamma$-line signals. 
In Fig.~\ref{fig:abcde} we show for the predictions A-E  what are the corresponding upper bounds on their decay rates $\Gamma_\gamma$ into monochromatic $\gamma$-rays, imposing that their continuum-part of CRs are not overproduced.
The strongest indirect bounds we get on the $\gamma$-ray lines are from the \pamela anti-proton data \cite{Adriani:2010rc}   (dashed lines)  or the \fermi isotropic $\gamma$-ray flux measurement \cite{Abdo:2010nz,FERMIsymposium} (solid lines), depending on the value of $m_{DM}$. 

The limits are obtained by taking $\Delta \chi^2 = \sum_i (\mathrm{data}_i-\mathrm{model}_i)^2/\sigma_i^2=9$. Here  ``model$_i$'' refers to the sum of the DM signal and our background flux model, and the standard deviations $\sigma_i$ are taken directly from \cite{Adriani:2010rc} and \cite{FERMIsymposium}.
The anti-proton flux can be well described by conventional astrophysics, and above 10 GeV (where solar modulation has a small effect) the canonical setup has relatively small uncertainties in its prediction. We therefore use only data above 10 GeV, and take our anti-proton background-flux model to exactly match the observation in each energy bin ($\chi^2=0$ by construction).
The origin of the isotropic diffuse $\gamma$-ray background is still quite uncertain and we thus take that background flux to be a completely free (positive definite) function of the $\gamma$-ray energy.\footnote{In practice, this means we only consider the $\gamma$ flux from DM in bins where it overshoots the observed flux when calculating $\Delta\chi^2$ (as $\chi^2=0$ by construction in the remaining bins).}
For the DM signal we take a NFW DM distribution (as in \cite{Cirelli:2010xx}) and include the extra Galactic contribution using the optical depth from Ref.~\cite{Gilmore:2009zb}. The Galactic signal is not fully isotropic, and we make the reasonable choice to consider the DM signal from the Galactic poles ($|b|\sim 90^\circ$).
For the anti-proton flux we choose the MIN propagation model (defined in \cite{Donato:2003xg,Cirelli:2010xx}) to always stay as conservative as possible in our below statements. For the MED and MAX propagation models (defined in \cite{Donato:2003xg,Cirelli:2010xx}) the various A-F curves in this figure would be approximately suppressed by a factor of 4 and 10, respectively. As Fig.~\ref{fig:abcde} shows, between the A and E cases the $\gamma$-line constraint varies by as much as a factor $\sim50$.

In Fig.~\ref{fig:abcde} these indirect upper bounds  on $\Gamma_\gamma$ can be compared to the current direct exclusion bounds on $\gamma$-ray line signals (grey shaded areas). Up to $E_\gamma=200$~GeV we used the limits from the $\gamma$-line searches by the \fermi collaboration \cite{Ackermann:2012qk}. From 200 GeV to 500 GeV we used the same $\Delta\chi^2 =9$ procedure as described above, applied to a $\gamma$-line on top of the isotropic diffuse $\gamma$-ray measurement \cite{FERMIsymposium}. Above $E_\gamma= 500$~GeV and up to 25 TeV the line signal constraints from \hess \cite{Abramowski:2013ax} are translated to limits on decaying DM. When using the \hess collaboration $\gamma$-line flux limits, we derived constraints by assuming a NFW DM profile (the limits are similar for an Einasto profile). We studied both their isotropic sky and central Galactic halo region data, but retain only the strongest constraints.

From Fig.~\ref{fig:abcde} one can directly see what could be the implications of a detection in the near future of a $\gamma$-line, of a CR excess or of both. To this end it should be kept in mind that if the operator considered carries a ``x'', ``m'' or ``v'' label, depending on what are the exact fields considered in this operator, its corresponding bound can be suppressed by  factors discussed above  compared to the maximal ratios $R_{\gamma/CR}$ shown in Fig.~\ref{fig:abcde}.

First of all, as Fig.~\ref{fig:abcde} shows, the A-E ratios turn out to have $\gamma$-ray line upper bounds from their CR constraints which are weaker than the present  direct observational limits on $\gamma$-lines (in a mild way for E though). This has the important implication that, with today's CR constraints, any of the operators could accommodate a possible near future situation where a $\gamma$-line would be observed without any observation of a CR excess.  
Only special cases could be excluded along such a scenario; basically the ``v'' labelled operators for given configurations such as the F$_Y$ and F$_L$ upper bounds  shown in Fig.~\ref{fig:abcde} (and marginally also some special ``x'' or ``m'' suppressed C-E cases).
Conversely, a broader energy continuum of photons, just below current constraints, could not be due to operators with the A to D  predictions. Similarly, the observation of an anti-proton excess (using in this case the ``MAX'' propagation parameters to stay conservative) could not be due to operators with predictions A to B. Obviously, an anti-proton (gamma continuum) excess can neither  be attributed to an operator's prediction if the DM particle mass is in a range above (below) $\sim 1$~TeV where the $\gamma$-continuum (anti-proton) constraints are stronger. 

This pattern could change in the future depending on with which precision  a CR excess and/or a $\gamma$ line would be observed or bounded. With the upcoming \amstwo data \cite{ams02ref}, and the potential to restrict the charged propagation model to be closer to the MED or MAX propagation models, the CR continuum constraint could potentially be improved by more than an order of  magnitude for $m_{DM}$ below a few TeV  \cite{Cirelli:2013hv}. In this case the A-E lines would lower accordingly. Similarly, Gamma-400 \cite{Galper:2012ji} and/or the CTA telescope \cite{Bringmann:2011ye,Bergstrom:2012vd,Doro:2012xx} could improve the $\gamma$-line sensitivities by one order of magnitude depending on the energy range and sky region (see \cite{Garny:2010eg} for the case of decaying DM).

For instance if a $\gamma$ line is observed, then the ratio of the observed $\gamma$-line decay rate and the maximally allowed $\gamma$-line decay rate, as given by the (updated) A-E predictions of Fig.~\ref{fig:abcde}, will tell us which operators can be excluded and which one could lead to an observation of an associated  CR excess. If this ratio is larger than 1, then the operator is excluded (and of course even more excluded if there are ``x'' or ``m'' suppressions, as this would increase the CR signal for an operator). If it is smaller than one, but not much smaller than one, then the CR excess associated to this $\gamma$-line is expected to be detectable.

Put in another way, taking into account these possible future experimental improvements, the following picture emerges. On the one hand, for the operators that involve the $F_Y^{\mu\nu}$ field strength, A and B cases, and if they have no ``m'' or ``x'' or ``v'' possibility of suppressions, it appears clearly improbable that in a near future we could see a CR excess associated to a $\gamma$ line. For A cases with index ``m'' and/or ``x'' this is not impossible, but difficult. For the A case with ``v'' label,  it is possible to see both associated signals easily, even without invoking any ``m'' or ``x'' suppressions (as in the F$_Y$ example). On the other hand for the operators that have a $F^{\mu\nu}_L$ field strengths, C-E cases, the discussion is the same but with all upper bounds shifted downward, so that it is easier to see both associated signals.

\section{Standard Model final state results}\label{sec:SM}

We now focus our analysis on the special case where all particles beside the DM one would be SM fields. This turns out to be a possibility for all operators (with $\psi$ replaced by the $e$, $\mu$ or $\tau$ doublet and the scalar $\phi$ fields replaced
by the SM scalar doublet $H\equiv(H^0,H^-)^T$ or its conjugate), with the exception of Eqs.~(\ref{OVDM5Y}), (\ref{OVDM5L}), (\ref{OphiDMYY'}), (\ref{OphiDMLY'}) and  (\ref{OV1DMY}). Note that there is no dimension 5 operator of this type for vector  DM particles, and for operators of dimension up to  6 the scalar DM particles can only have the quantum numbers (1/3/5,0), (2/4/6,1), (4/6,3) or (6,5). For fermion DM particles only  the (1/3/5,0), (3/5,2) or (2/4,1), (4,3) multiplets are possible.
The values of $R_{\gamma/CR}$ when the operators include only SM decay products are particularly interesting because they are almost totally fixed.
Indeed, in these cases the $R_{\gamma/CR}$ ratios  neither depend on any unknown final state that might or might not give CRs, nor on the unknown vev  of new fields.  For a given operator the $R_{\gamma/CR}$ ratio may only depend on the way the $SU(2)_L$ indices are contracted. 
Here we will consider only the case where the DM particle has vanishing hypercharge ({\it e.g.}~to avoid constraints on direct detection processes mediated by a $Z$ boson) and the so called ``inert doublet'' candidate where the DM multiplet is a scalar doublet with the same hypercharge as the SM scalar doublet.

\begin{figure}[t!]
\center{\includegraphics[width=0.95 \columnwidth]{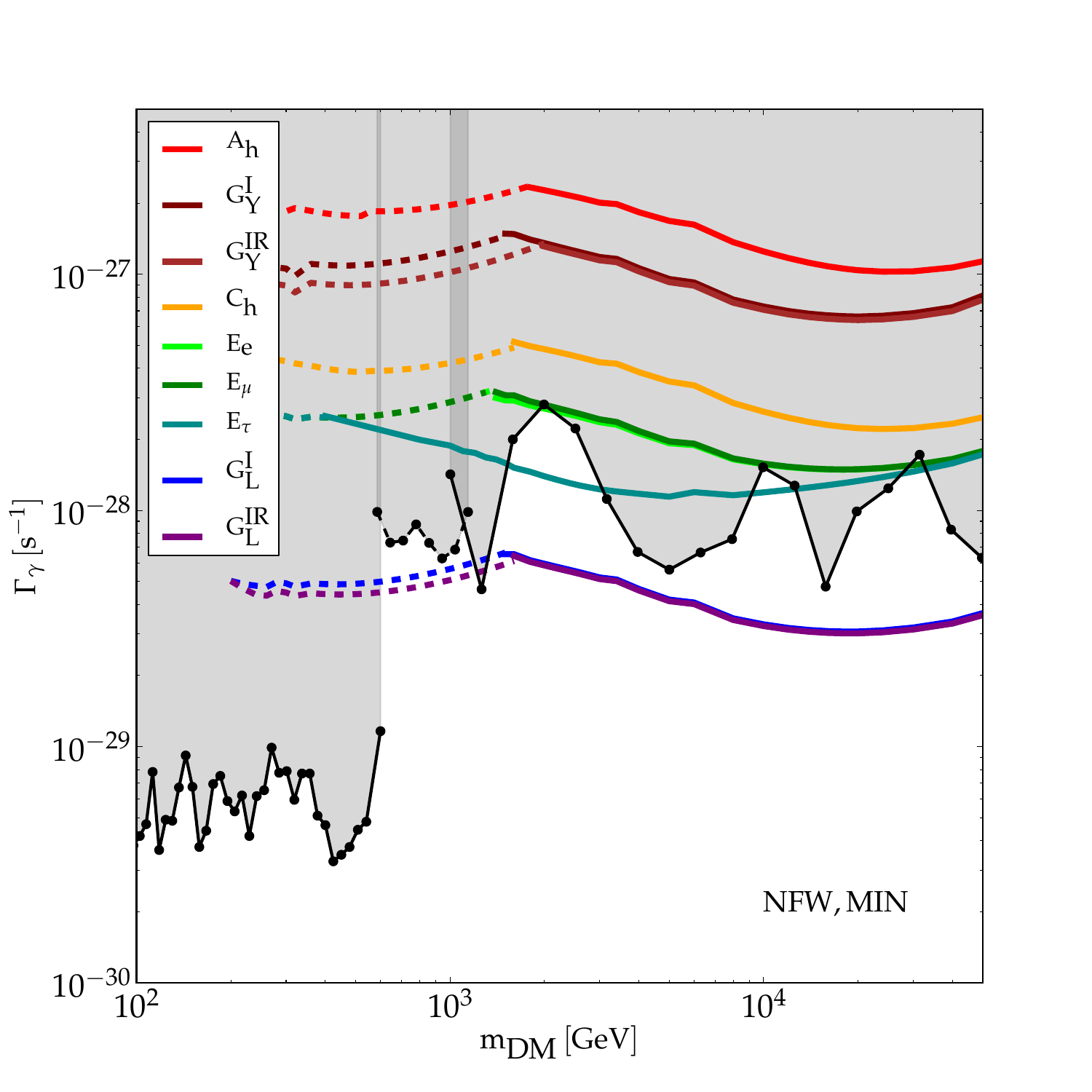}}
\caption{Same as for Fig.~\ref{fig:abcde} but for SM final states. Note that only the $ E_{e, \mu, \tau} $ and not the $ D_{e, \mu, \tau} $ predictions have been plotted for readability sake, but their ratios are the same as the $E$ to $D$ ratio that can be found in Fig.~\ref{fig:abcde}. The curves are, from top to bottom, in the same order as mentioned in the figure's legend. } \label{fig:SM}
\end{figure}

Depending on the operator, the hyperchargeless candidates lead to predictions A-E or new predictions, which we call $D_{e,\mu,\tau}$, $E_{e,\mu,\tau}$, $A_h$ and $C_h$. The predictions $D_{e,\mu,\tau}$ and $E_{e,\mu,\tau}$  correspond to the cases where a $e$, $\mu$ or $\tau$ charged lepton is produced in association to a $W$ in the final state. They are obtained from the  $D$ and $E$, respectively, by simply replacing $n_{CR/W^\pm}$ by $n_{CR/W^\pm}+n_{CR/l^\mp}$ in Eq.~(\ref{ngammanZphi1LL}). The predictions $A_h$ and $C_h$, corresponding to when a SM scalar boson is produced with the $\gamma$ and $Z$ in the 2-body final state, are obtained by simply adding a $n_{CR/h}$ term in the denominator of the prediction $A$ and $C$, respectively. The resulting constraints are given in Fig.~\ref{fig:SM}, and the dictionary between signal predictions and possible effective operators for these hyperchargeless multiplets is given in Table~\ref{tab:1}.

For the inert scalar doublet case one can get A, B, E or new lengthy predictions \cite{GHTfuture} from the "v" labelled operators  in Eqs.~(\ref{OphiDMY})-(\ref{OphiDML}), which we call $G_{Y}^{I/IR}$ and $G_{L}^{I/IR}$ and depend on whether the DM particle is made of the real (R), of the imaginary (I), or of both components (RI) of the neutral component of the inert doublet (taking the SM scalar vev to be real by convention). Depending on whether the $\lambda_5$ term in the tree-level scalar potential gives a positive, negative or vanishing mass splitting between the two neutral components (see {\it e.g.}~\cite{Barbieri:2006dq}), these $G^{I/IR}$ predictions are different because, as for the $\lambda_5$ term, these two operators break the Peccei-Quinn type symmetry between these two components.
The signal to operator dictionary for this case is given in Table~\ref{tab:2}.

\begin{table}
\begin{tabular}{@{\extracolsep{\fill}} l|l|l} 
   Signal $R_{\gamma/CR}$      		&  Operator in Eq.		& Quantum no.	 $(n,Y)$		\\ \hline\hline 
    A				&  (1)	 		& $\phi_{DM}$ =(1,0)  						\\
    			    &  (17)			& $\psi_{DM}$=(1,0), \, (3,0)  					\\ \hline
     $A_h$			& (24),(26)		& $V_{DM}=(1,0)$,\,  $\bar{H} H =(1,0)$ \\   \hline
    B 				&  (2)			& $\phi_{DM}$=(1,0),\, (3,0)  					\\ \hline
    C	& (18)				& $\psi_{DM}$ =(3,0), \,$\bar{\psi}_lH=(1,0)$ 		\\ \hline
    $C_h$			& (25),(27)		& $V_{DM}=(1,0)$,\,  $\bar{H} H =(3,0)$\\   \hline
    D   			&  (18)			& $\psi_{DM}$ =(5,0)	 	  				\\  \hline
    $D_l$		& (18)				& $\psi_{DM}$ =(3,0), \,$\psi_{DM} H=(2/4,0)$ 		\\ \hline
    E			&  (18)			& $\psi_{DM}$ =(1,0)	 	  				\\  \hline
    $E_l$  	& 	(18)			& $\psi_{DM}$ =(3,0),  \,  $\bar{\psi}_l\psi_{DM}=(2/4,1)$ 	\\      \hline
 no-lines	&  (3)			& $\phi_{DM}$=(3,0)   	  				\\
    		& (18)				& $\psi_{DM}$ =(3,0), \,$\bar{\psi}_lH=(1/3,0)$ 		\\ \hline
    \hline
\end{tabular}
\caption{All possible signal scenarios for a hyperchargeless DM candidates decaying into SM particles. The labels of the monochromatic-line to cosmic-ray signal ratios $R_{\gamma/CR}$ are explained in Eq.~(\ref{ngammanZmax}) - (\ref{ngammanZphi1LL}) and in the text of Sec.~\ref{sec:SM}.} \label{tab:1}
\end{table}
 
\begin{table}
    \begin{tabular}{@{\extracolsep{\fill}} l|l|l} 
Signal $R_{\gamma/CR}$     &  Operator in Eq.		& Quantum no.	 $(n,Y)$    \\ \hline\hline 
    A 			&  (\ref{OphiDMYY}) 				& --  					\\ \hline
    B 			&  (\ref{OphiDMYL})				& -- 					\\ \hline
    E 			& 		 (\ref{OphiDMLL}) 					&  $F_L F_L = (1/5,0)$, $\phi_{DM}F_L = (2/4,-1)$ \\ \hline
         $G_Y^{I/IR}$ 	& 	{	 (\ref{OphiDMY})		}	& DM\,$\equiv$ Im$\left[\Phi^0_{DM}\right]$/ Re+Im$\left[\Phi^0_{DM}\right]$		\\ \hline
         $G_L^{I/IR}$ 	& {		(\ref{OphiDML})	}		&  DM\,$\equiv$  Im$\left[\Phi^0_{DM}\right]$/ Re+Im$\left[\Phi^0_{DM}\right]$\\ \hline
    no-lines		&		 (\ref{OphiDMLL}) 					& $F_L F_L=(3,0)$ \\ 
	& 		 (\ref{OphiDMY}), (\ref{OphiDML})					&  DM\,$\equiv$ Re$\left[\Phi^0_{DM} \right]	$			\\ 
        \hline\hline

      \end{tabular}
    \caption{Similarly to Table I, all possible signal scenarios for an inert scalar doublet \mbox{$\phi_{DM} \equiv (2,1)$} decaying into SM particles.}
    \label{tab:2}
\end{table}

Fig.~\ref{fig:SM} shows that, similarly to what one observed in Fig.~\ref{fig:abcde}, almost all cases with only SM final states have potential $\gamma$-ray line signal strengths that are not restricted by their associated continuum of CRs. The only exception where actual direct limit on a $\gamma$-line  are weaker is for the $G_{L}$ cases --  mainly because the photon-line emission for this ``v'' labelled operator in Eq.~(\ref{OphiDML}), is suppressed by a factor of $T_3^2=1/4$ with respect to the $W^+ W^-$ emission. 

Note that for the cases with prompt charged leptons in the final state, leading to $(D/E)_{e,\mu, \tau}$ predictions shown in Fig.~\ref{fig:SM},  we have checked that the $\gamma$ flux from inverse Compton interactions of these leptons can be neglected. As for the observational positron and electron constraints \cite{Adriani:2008zr, Aharonian:2008aa,Ackermann:2010ij}, they turn out to not  give any further bounds here.

Note finally that in Table.~1 there is a case where the DM field is a quintuplet. Assuming that there is no beyond the SM fields other than this quintuplet, this is the quintuplet of the ``minimal DM" setup \cite{Cirelli:2005uq}. The neutral component of this quintuplet is naturally stable enough because "accidentally" it cannot appear in any operator of dimension less than 6. Table.~1 shows that it can lead to a $\gamma$-line through the operator of Eq.~(\ref{Opsi1DML}),
leading to prediction $E_{e,\mu,\tau}$ depending on which SM lepton doublet is considered in the operator, see Fig.~\ref{fig:SM}.

\section{A few final comments}
\label{sec:final comments}

In Figs.~\ref{fig:abcde} and \ref{fig:SM} we took into account only the full list of two-body decays the various operators give.
This is expected to be a good approximation, except for the following case: if the two-body decays proceed through the vev of a scalar and if the DM mass is far above the value of this vev.  In this case, if the three-body decay is kinematically allowed, the two to three-body decay-width ratio is expected to be of order $\sim 48 \pi^2 \langle \phi \rangle^2/m_{DM}^2$ . 
Finally note that a UV theory is of course not necessarily expected to induce only one relevant operator. In presence of several operators the CR production could be easily boosted relatively to the $\gamma$-line signal, so that the A to G $\gamma$-line bounds in Figs.~\ref{fig:abcde} and \ref{fig:SM} would lower accordingly. Alternatively, with several operators there exists also the possibility that the UV physics induces several effective operators that would produce the same final states, leading to a destructive or constructive interference for some of the channels. This issue is in many respects similar to the well-known one of determining the cut-off of possible new physics  that would induce operators composed of only SM fields
\cite{Buchmuller:1985jz} ({\it i.e.}\  lower bounds on UV  cut-offs given in the literature are typically derived for single operators, and they could be decreased if interferences between operators are allowed to be tuned).  
The variation in $R_{\gamma/CR}$ that such interferences could cause does highly depend on the operators considered and the multiplet considered in them. For many cases it is not possible to increase sizably $R_{\gamma/CR}$ in this way. For other cases it is possible by invoking one or several tunings of the operators' relative contributions ({\it i.e.}~of couplings and quantum numbers in the UV theory). 

In a similar vein one has also to keep in mind that, through renormalization group effects, a single operator generated at a high energy scale can generate other ones at low scale (although in general with sub-leading coefficients).

\medskip
In conclusion only a few dim $\leq 6$ effective operator structures can lead to a monochromatic $\gamma$-line from DM decay, even if one accounts for possible new BSM particles in the decay products. Each of these operators leads to a $\gamma$-line to CR flux-ratio which turns out to be bounded from above and in almost all cases also from below. This leads to clear possibilities of observing a continuum of CRs associated to a detected $\gamma$-line. From the observation of a $\gamma$-line and/or of a CR fluxes that these operators could induce, and from the observation/non-observation of multiple $\gamma$-lines (with characteristic energy and height), there exists a clear potential to constrain/discriminate various sets of operators.

\vspace{5mm}
\section*{Acknowledgement}
\vspace{-3mm}
We acknowledge stimulating discussions with X.~Chu, M.~Cirelli, M.~Pshirkov, M.~Tytgat and G.~Vertongen.
This work is supported by the FNRS-FRS, the IISN, the ULB-ARC grant project `Fundamental 
aspects of gravitational interactions', and the Belgian Science Policy, IAP VII/37.

\end{document}